\documentclass[twocolumn,secnumarabic,amssymb, nobibnotes, aps, prb,reprint]{revtex4-1}

\usepackage{ulem}
\newcommand{\rsq}{\mathrm{R}_\mathrm{sq}}

\newcommand{\bpar}{\mathrm{B}_\parallel}

\newcommand{\bper}{\mathrm{B}_\perp}

\newcommand{\rnl}{\mathrm{R}_\mathrm{NL}}
\newcommand{\rnlz}{\mathrm{R}_\mathrm{NL0}}

\newcommand{\ds}{\mathrm{D}_{\mathrm{s}}}

\newcommand{\ts}{\tau_{\mathrm{s}}}

\newcommand{\ohm}{\Omega}
\newcommand{\tpar}{\tau_\parallel}

\newcommand{\rc}{\mathrm{R}_\mathrm{c}}

\newcommand{\Pa}{\mathrm{P}_\mathrm{A}}
\newcommand{\Pb}{\mathrm{P}_\mathrm{B}}

\newcommand{\idc}{\mathrm{I}_{\mathrm{DC}}}
\newcommand{\pind}{\mathrm{p}_\mathrm{in}}
\newcommand{\pdd}{\mathrm{p}_\mathrm{d}}

\usepackage{graphicx}% Include figure files
\usepackage{color}
\usepackage{textcomp}
\usepackage{amsmath}
\usepackage{dcolumn}% Align table columns on decimal point
\usepackage{bm}% bold math

\begin{document}

\title{Efficient spin injection into graphene through trilayer hBN tunnel barriers}%

\author{Johannes Christian Leutenantsmeyer}%
\email[]{These authors contributed equally to this work, \\E-Mail: j.c.leutenantsmeyer@rug.nl}
\affiliation{Physics of Nanodevices, Zernike Institute for Advanced Materials, University of Groningen, 9747 AG Groningen, The Netherlands}

\author{Josep Ingla-Ayn\'{e}s$^*$}%
\affiliation{Physics of Nanodevices, Zernike Institute for Advanced Materials, University of Groningen, 9747 AG Groningen, The Netherlands}

\author{Mallikajurna Gurram}%
\affiliation{Physics of Nanodevices, Zernike Institute for Advanced Materials, University of Groningen, 9747 AG Groningen, The Netherlands}

%\author{A.A. Kaverzin}%
%\affiliation{Physics of Nanodevices, Zernike Institute for Advanced Materials, University of Groningen, 9747 AG Groningen, The Netherlands}

\author{Bart J. van Wees}%
\affiliation{Physics of Nanodevices, Zernike Institute for Advanced Materials, University of Groningen, 9747 AG Groningen, The Netherlands}

\begin{abstract}
We characterize the spin injection into bilayer graphene fully encapsulated in hBN using trilayer (3L) hexagonal boron nitride (hBN) tunnel barriers. As a function of the DC bias, the differential spin injection polarization is found to rise up to -60\% at -250~mV DC bias voltage. We measure a DC spin polarization of $\sim$ 50\%, a 30\% increase compared to 2L-hBN. The large polarization is confirmed by local, two terminal spin transport measurements up to room temperature. 
We observe comparable differential spin injection efficiencies from Co/2L-hBN and Co/3L-hBN into graphene and conclude that possible exchange interaction between cobalt and graphene is likely not the origin of the bias dependence. 
Furthermore, our results show that local gating, arising from the applied DC bias is not responsible for the DC bias dependence. Carrier density dependent measurements of the spin injection efficiency are discussed, where we find no significant modulation of the differential spin injection polarization. We also address the bias dependence of the injection of in-plane and out-of-plane spins and conclude that the spin injection polarization is isotropic and does not depend on the applied bias. 

%Can we say that it seems like a bandstructure effect? I wouldn't
%Our findings exclude local gating by the hBN bias and proximity effects between cobalt and hBN as origin for the large spin injection efficiencies. 

%We can claim:
%\begin{itemize}
%	\item tunability for TL- as for BL-hBN, which is very relevant since it excludes a bunch of potential origins
%	\item high spin polarization at zero bias (20\% with TL vs. $<$5\% for BL)
%	\item carrier concentration does not change the dependence on the bias
%	\item DC-2T measurements of spin valve and Hanle
%	\item Dependence of the spin transport parameters on the barrier resistance
%\end{itemize}

\end{abstract}

\date{\today}%
\maketitle

\section{Introduction}
Graphene is an ideal material for long distance spin transport experiments due to its low intrinsic spin-orbit coupling and outstanding electronic quality \cite{Huertas-Hernando2006,Han2014,Roche2015,Ingla-Aynes2016, Drogeler2016}.
Experimental results have shown that long spin relaxation lengths require the protection of the graphene channel from contamination \citep{Ingla-Aynes2016, Drogeler2016,Zomer2012,Guimaraes2014}. The most effective way to achieve this is the encapsulation of graphene with hexagonal Boron Nitride (hBN), which substantially improved the spin transport properties \cite{Zomer2012,Guimaraes2014,Drogeler2014,Ingla-Aynes2015,Drogeler2016,Gurram2016,Singh2016}. Besides of the cleanliness of the channel, the efficient injection and detection of spins into graphene is an essential requirement to fabricate high performance devices. To circumvent the conductivity mismatch problem \cite{Schmidt2000}, a tunnel barrier is employed to enhance the spin injection polarization \cite{Rashba2000}. While commonly used Al$_2$O$_3$ and TiO$_2$ tunnel barriers yield typically spin polarizations below 10\% \cite{Jozsa2009}. The use of crystalline MgO \cite{Han2010,Volmer2013,Volmer2014}, hBN \cite{Kamalakar2015a,Kamalakar2016,Gurram2017}, amorphous carbon \cite{Neumann2013} or SrO \cite{Singh2017SrO} as tunnel barrier has led to significant enhancements. In particular, the use of a 2L-hBN flake for spin injection gives rise to bias dependent differential spin injection polarizations p up to p = 70\%, which is defined as the injected AC spin current i$_\mathrm{s}$ divided by the AC charge current i$_\mathrm{AC}$. Furthermore, 2L-hBN provides contact resistances in the range of 10~k$\Omega$, which can be close to the spin resistance of high quality graphene and affect spin transport \cite{Gurram2017}. 3L-hBN tunnel barriers promise higher contact resistances, leaving the spin transport in 3L-hBN/graphene unaffected \cite{Kamalakar2016,Gurram2018}.

While the underlying mechanism for the DC bias dependent spin injection is still unclear, ab initio calculations of cobalt separated from graphene by hBN show, that in the optimal case Co can induce an exchange interaction of 10~meV even through 2L-hBN into graphene \cite{Zollner2016}, therefore, a comparison between hBN tunnel barriers of different thicknesses can give insight on the proximity effects between graphene and cobalt. 

Here we show that 3L-hBN tunnel barriers increase the differential spin injection polarization into bilayer graphene (BLG) from a zero bias value of p = 20\% up to values above p = -60\% at negative DC bias. The DC spin injection polarization P, which is defined as the DC spin current I$_\mathrm{s}$ divided by the DC charge current I$_\mathrm{DC}$, increases up to P = 50\%, at a DC bias current of -2~$\mu$A. This is a substantial advantage over 2L-hBN, which shows P $\sim$ 35\%. The large DC spin polarization allows us to measure spin signals in a DC two terminal spin valve geometry up to room temperature. %and spin precession at room temperature. 
We show that the differential spin injection polarization is, contrary to Ringer et al. \cite{Ringer2018}, independent of the carrier density. The rotation of the magnetization of the electrodes out-of-plane under a perpendicular magnetic field $\bper$ allows us to study the bias dependence of the spin injection polarization of out-of-plane spins ($\mathrm{p_z}$). We compare $\mathrm{p_z}$ with the in-plane polarization $\mathrm{p_y}$ and conclude that $\mathrm{\mathrm{p_z}/\mathrm{p_y}}$ $\sim$ 1, independently of the applied DC bias. 

\section{Sample preparation and contact characterization}

\begin{figure}[tb]
\centerline{\includegraphics[width=\linewidth]{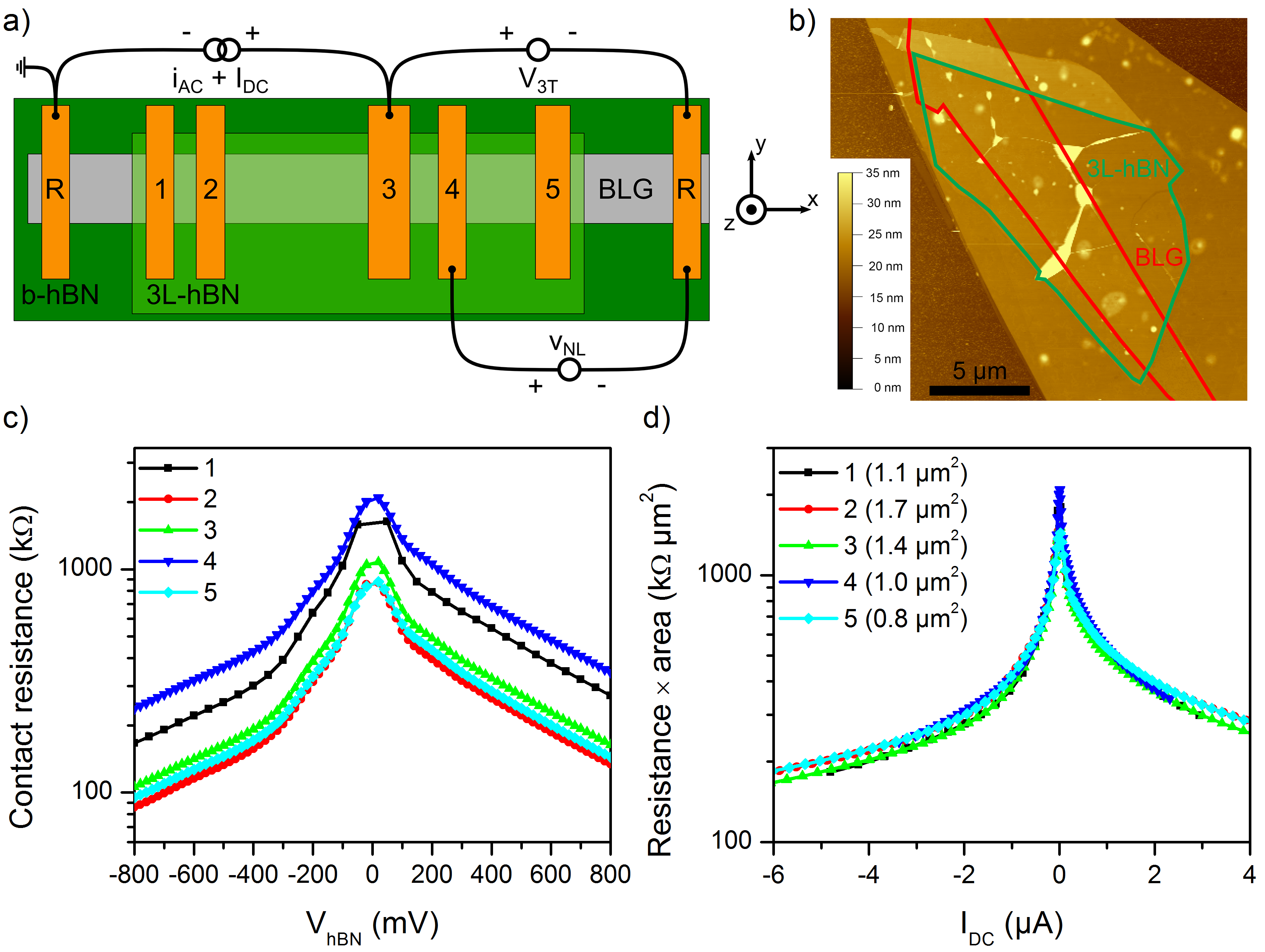}}
\caption{a) Schematic device geometry. A BLG flake is encapsulated between a 5~nm thick hBN (b-hBN) and a 1.2~nm 3L-hBN flake, used as a tunnel barrier for spin injection. The different measurement geometries are sketched. V$_\mathrm{3T}$ is the DC voltage across the hBN tunnel barrier, from which the contact resistance can be calculated via R$_\mathrm{c}$ = V$_\mathrm{3T}$/I$_\mathrm{DC}$. v$_\mathrm{NL}$ is the AC non-local voltage and used to calculate the non-local resistance $\rnl$ =  v$_\mathrm{NL}$/i$_\mathrm{AC}$. Additionally to the AC measurement current i$_\mathrm{AC}$, a DC current I$_\mathrm{DC}$ is applied to bias the injector contact. Note that the outer reference contacts (R) do not have an hBN tunnel barrier. b) Atomic force microscopy image of the hBN/BLG/3L-hBN heterostructure before the contact deposition. c) Contact resistance measurements for different voltages applied across the hBN tunnel barrier (V$_\mathrm{3T}$). d) The calculated resistance-area products ($\rc\times\mathrm{A}$) range between 180~k$\ohm\mu$m$^2$ and 2~M$\ohm\mu$m$^2$, depending on the applied DC bias current I$_\mathrm{DC}$.
\label{Figure1TL}%C1=13, C2=11, C3=9, C5=39, C4=7
}
\end{figure}

The device geometry is shown in Fig.~\ref{Figure1TL}a. BLG is encapsulated between a 5~nm thick bottom hBN and a 1.2~nm thick 3L-hBN flake, which acts as a tunnel barrier. The stack is deposited on a silicon oxide substrate with 90~nm oxide thickness, that is used to tune the carrier concentration in the graphene channel. This device has been used to study the spin lifetime anisotropy in BLG \cite{Leutenantsmeyer2018}.  Unless noted, all measurements are carried out at T = 75~K to improve the signal to noise ratio. The atomic force microscopy image of the stack before the contact deposition is shown in Fig.~\ref{Figure1TL}b. The contact resistances are characterized by measuring the bias dependence in the three terminal geometry, R$_\mathrm{c}$ = V$_\mathrm{3T}$/I$_\mathrm{DC}$, and shown in Fig.~\ref{Figure1TL}c as a function of the voltage applied across the 3L-hBN tunnel barrier (V$_\mathrm{3T}$). The bias dependent contact resistances are normalized to the contact area and plotted as a function of the DC current I$_\mathrm{DC}$ applied to the hBN barrier in Fig.~\ref{Figure1TL}d. To determine the spin transport properties of our device, we use the standard non-local geometry \cite{Jedema2001,Jedema2002,Tombros2007}, the circuit is shown in Fig.~\ref{Figure1TL}a. An AC charge current i$_\mathrm{AC}$ is applied together with I$_\mathrm{DC}$ between the injector and the left reference contact, which does not have any tunnel barrier and therefore does not inject spins efficiently. Because of the spin polarization of the cobalt/hBN contacts, the injected charge current is spin polarized and induces a spin accumulation into the channel. The spins diffuse in the BLG channel and are detected by a second cobalt/hBN contact in the non-local geometry. 

\section{Spin transport at different DC bias currents}
\begin{figure}[tb]
\centerline{\includegraphics[width=\linewidth]{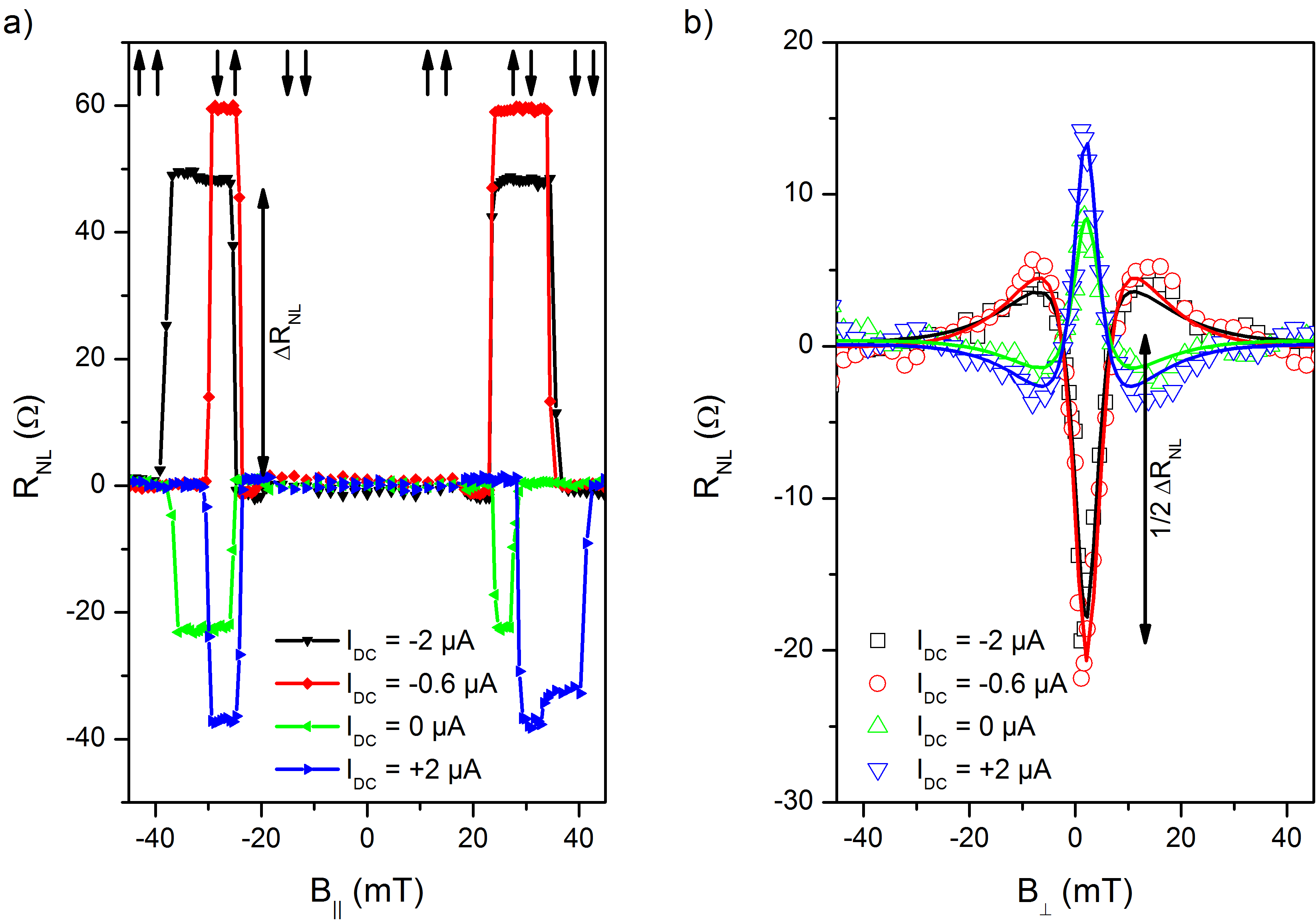}}
\caption{Characterization of the spin transport in the fully hBN encapsulated BLG device at different DC bias currents using Contact~1 as injector and Contact~5 as detector. Both electrodes are separated by L = 10~$\mu$m. a) Non-local resistance $\rnl$ measured in an in-plane magnetic field $\bpar$ where the magnetization of the injector and detector contacts are switched between parallel and antiparallel alignment. b) Spin precession measurement in an out-of-plane magnetic field $\bper$. The fitting using the Bloch equations yields the spin transport parameters shown in Table~\ref{Table1_TL}. Note that non-local background resistances smaller than 35~$\Omega$ have been subtracted from the data to compare the influence of the different DC bias.%13-39 C1=13, C2=11, C3=9, C4=7, C5=39
\label{Figure2TL}}
\end{figure}

The different coercive fields of the cobalt contacts allow the separate switching of individual electrodes with an in-plane magnetic field $\bpar$ and the measurement of the non-local resistance ($\rnl$ = v$_\mathrm{NL}$/i$_\mathrm{AC}$) in different magnetic configurations. The non-local spin valve is shown in Fig.~\ref{Figure2TL}a for different DC bias currents. The abrupt signal changes are caused by the switching of the contact magnetization, the magnetization configurations are indicated with arrows. The spin signal $\rnl$ is determined by the difference between parallel ($\rnl$($\uparrow \uparrow$) = $\rnl$($\downarrow \downarrow)$) and antiparallel ($\rnl$($\uparrow \downarrow$)  = $\rnl$($\downarrow \uparrow$)) configurations.

The most accurate way to characterize the spin transport properties of the channel is using spin precession, where the magnetic field is applied perpendicular to the BLG plane ($\bper$), causing spins to precess in the x-y plane. By fitting $\rnl$ to the Bloch spin diffusion equations, we extract the spin lifetime ($\ts$), spin diffusion coefficient ($\ds$) and the average polarization of both electrodes ($\mathrm{p_y}$). The data is shown for different DC bias currents in Fig.~\ref{Figure2TL}b, the fitting curves are shown as solid lines. Note that the spin transport parameters in Table~\ref{Table1_TL} are within the experimental uncertainty for all $\mathrm{I_{DC}}$ values. 
Therefore, we average $\ts$, $\ds$, and the spin relaxation length ($\lambda$) over all four values and obtain $\ts$ = (1.9	$\pm$ 0.2)~ns, $\ds$ = (183	$\pm$ 17)~cm$^2$/s and $\lambda$ = $\sqrt{\ds \tpar}$ = (5.8 $\pm$ 0.6)~$\mu$m. These parameters are comparable to the ones reported in Ref.~\cite{Gurram2018}. 
%We observe that the spin relaxation length is also in good agreement with the value extracted from distance dependent spin valve measurements, discussed in the supplementary information. 
We conclude that the change in contact resistance with $\mathrm{I_{DC}}$ does not affect the spin transport for values above 100~k$\ohm$. This is caused by the fact that the contact resistance remains clearly above the spin resistance of the channel R$_\mathrm{s}$ = $\rsq\lambda$/w $\sim$ 1.8~k$\ohm$, where $\rsq$ is the graphene square resistance and w the graphene width \cite{Maassen2012}.

\begin{table}[bt]
\caption{Spin transport parameters extracted from the data shown in Fig.~\ref{Figure2TL}b. The values obtained from averaging over the different $\mathrm{I_{DC}}$ are: $\ds$ = (183 $\pm$ 17)$~\mathrm{cm^2/s}$, $\ts$ = (1.9 $\pm$ 0.2)~ns and $\lambda$ = (5.8 $\pm$ 0.6)~$\mu$m.}
\begin{ruledtabular}
\begin{tabular}{c c c c c}
$\mathrm{I_{DC}}$ & $\rc\times\mathrm{A}$ & $\ds$& $\ts$& $\lambda$ \\
($\mu$A) & (k$\ohm\mu$m$^2$) &($\mathrm{cm^2/s}$) & (ns) & ($\mu$m)  \\
\hline
-2 & 280 & 208 $\pm$	25 & 2.1 $\pm$ 0.2 & 6.4 $\pm$ 1.6 \\
-0.6 & 760 & 177	$\pm$ 21 & 1.7 $\pm$ 0.2 & 5.5 $\pm$	1.2 \\
0 & 2100 & 171 $\pm$ 24 & 1.7 $\pm$ 0.2 & 5.4 $\pm$ 1.5 \\
+2 & 380 & 177 $\pm$ 24 & 2.0 $\pm$	0.2 & 5.8 $\pm$ 1.5 \\
\end{tabular}
\end{ruledtabular}
\label{Table1_TL}
\end{table}
 
Note that the spin resistance of graphene can exceed 10~k$\ohm$ in high quality devices. This is close to the contact resistance of biased 2L-hBN tunnel barriers, which typically range, depending I$_\mathrm{DC}$, between 5~k$\ohm$ and 30~k$\ohm$ \cite{Leutenantsmeyer2018YIG}. Furthermore, the extended data sets discussed in the supplementary information and our analysis in Ref.~\cite{Leutenantsmeyer2018} confirm that contact induced spin backflow is not limiting spin transport for contact resistances above 100~k$\Omega$.

\section{DC bias dependence of the differential spin injection efficiency}
\begin{figure}[tbh]
\centerline{\includegraphics[width=\linewidth]{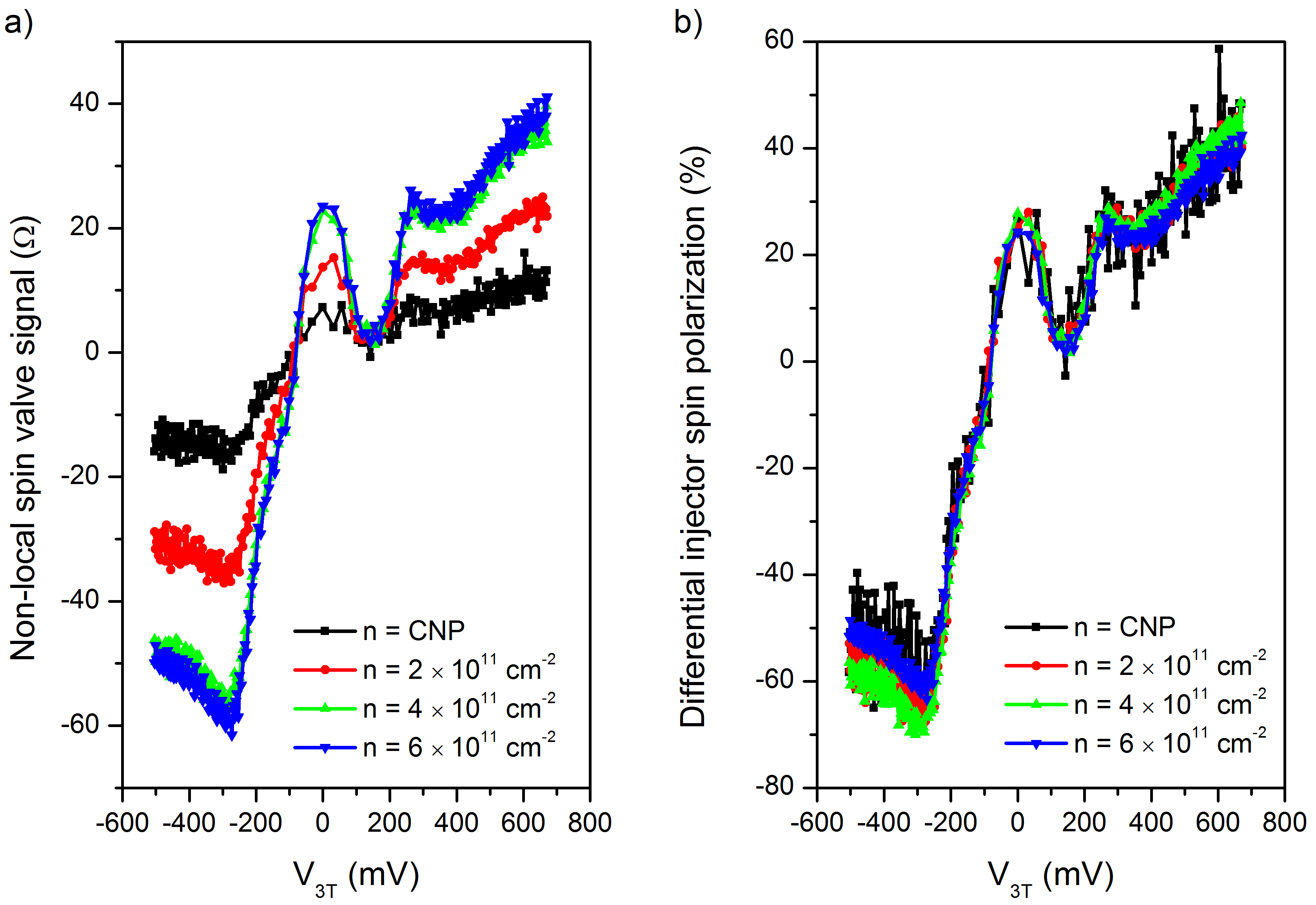}}
\caption{a) Measurement of the DC bias dependence of the $\rnl$ at four different carrier concentrations, where Contact~1 is used as injector and Contact~5 as detector. b) The extracted spin polarization of the injector contact using equation~\ref{Equation1TL}. The spin polarization reaches -60\% at negative and +40\% at positive I$_\mathrm{DC}$. Measurements using Contact~2 as injector yield comparable results.
\label{Figure3TL}
}
\end{figure}

In Fig.~\ref{Figure3TL}a we show the non-local spin valve signal $\Delta\rnl$ =$\rnl(\uparrow \uparrow)$ - $\rnl(\uparrow \downarrow)$. For a comparison with 2L-hBN tunnel barriers, we calculate V$_\mathrm{3T}$, the voltage applied to the tunnel barrier, by using the current-voltage characteristics of each contact. To resolve small features in the bias dependence, we use measurement currents as low as i$_\mathrm{AC}$ = 50~nA. As observed for 2L-hBN barriers \cite{Gurram2017,Leutenantsmeyer2018YIG}, $\Delta \rnl$ changes sign at V$_\mathrm{3T}$ $\sim$ -100~mV, which we also observe with a 3L-hBN barrier. 
Our data also shows additional features: Firstly, $\vert\Delta \rnl\vert$ shows a maximum at V$_\mathrm{3T}$ $\sim$ -250~mV and decreases again for V$_\mathrm{3T}$ $<$ -250~mV. In contrast, we observe a continuous increase for V$_\mathrm{3T}$ $>$ +300~mV. Secondly, we observe a peak at zero V$_\mathrm{3T}$, indicating that the polarization of Co/3L-hBN at zero DC bias is higher than in Co/2L-hBN. 
Note that 2L-hBN devices in Ref.~\cite{Leutenantsmeyer2018YIG} show also these small features around zero DC bias (Fig.~\ref{Figure3_5TL}b). 
%We believe that these features were not observed for bilayer hBN tunnel barriers due to the relatively large 3~$\mu$A AC measurement current used in the previous studies. \jcladd{Problem: that corresponds to 50 mV, which is probably not enough to smear out the feature. Still, we measured in terms of V farther than Mallik.}%by \textcite{Gurram2017}. 

To calculate the polarization of the Co/hBN interface from $\Delta \rnl$, we use:
\begin{equation}
\mathrm{\Delta \rnl=\frac{\pind \pdd \rsq\lambda}{w}e^{-d/\lambda}}\label{Equation1TL}
\end{equation}
where $\pind$ and $\pdd$ are the differential injector and detector spin polarizations, and $\mathrm{d}$ the separation between injector and detector. An overview of all extracted spin transport parameters is shown in the supplementary information. 
%To determine both $\mathrm{P_i}$ and $\mathrm{P_d}$, we have used an additional contact as detector ($\mathrm{d2}$) and measured simultaneously $\Delta \rnl$ at $\mathrm{d2}$. 
%We obtain $\mathrm{P_i\times P_d}$ and $\mathrm{P_i\times P_{d2}}$, which is a system with two equations and three unknown parameters. The last equation can be obtained from measuring $\Delta \rnl$ using the first detector $\mathrm{d}$ as injector and $\mathrm{d2}$ as a detector to get $\mathrm{P_d\times P_{d2}}$. \jcladd{not sure if this is not too much detail}
Following this procedure for I$_\mathrm{DC}$ = 0 at different configurations we obtain the unbiased spin polarizations of all contacts of p$_1$ = 24\%, p$_2$ = 23\%, p$_3$ = 30\%, p$_4$ = 36\%, and p$_5$= 38\%. Since $\pdd$ does not depend on the DC bias, which is applied to the injector only, we can calculate the bias dependence of $\pind$ (Fig.~\ref{Figure3TL}b). The absolute sign of p cannot be determined from spin transport measurements \cite{Gurram2017}, and we define p to be positive for I$_\mathrm{DC}$ = 0. 

Note that the slope observed in Fig.~\ref{Figure3TL}b is in qualitative agreement with the ab-initio calculations by Piquemal-Banci et al. \cite{Piquemal-Banci2018} for chemisorbed cobalt on hBN, suggesting that the observed DC bias dependence arises from the Co/hBN interface and not from proximity coupling between cobalt and graphene. 

We conclude that $\pind$(I$_\mathrm{DC}$) can reach values comparable to 2L-hBN tunnel barriers. Moreover, the comparison between different carrier concentrations shows that the spin injection polarization does not depend of the carrier density, even at the charge neutrality point. This also indicates that local spin drift in the barrier arising from pinholes is not responsible for the bias dependence. The drift velocity is inversely proportional to the carrier density, and therefore, the effect of spin drift is the largest near the neutrality point \cite{Ingla-Aynes2016}. Furthermore, if charge carrier drift in the channel would be relevant, the measured Hanle curves would widen \cite{Huang2008}. Consequently, the extracted spin lifetimes would decrease with increasing I$_\mathrm{DC}$, which we do not observe here. 
Furthermore, our I$_\mathrm{DC}$ is at most 2~$\mu$A, whereas a sizable drift effect requires larger charge currents \cite{Ingla-Aynes2016}. Local charge carrier drift at the injector, caused by pinholes in the barrier, was used to explain a modulation of the spin injection polarization \cite{Jozsa2009}. From our measurements we can exclude this mechanism as origin due to the negligible modulation of the spin injection polarization with n. Moreover, we use crystalline hBN as tunnel barrier, which has the advantage over evaporated barriers that pinholes are not expected to be present.

%Note, however that $\mathrm{P_i}$ and $\mathrm{P_d}$ are differential spin polarizations ($\mathrm{P=I_s^{AC}/I_{ac}}$ where $\mathrm{I_s^{AC}}$ is the injected AC spin current).\\

\section{Calculation of the DC spin polarization}
For practical applications, a large DC spin polarization P is required. Using the differential spin polarization $\mathrm{p}$, we can calculate P via \cite{Gurram2017}:
\begin{align}
\mathrm{p(I_{DC}) =\frac{d P(I_{DC})}{d I_{DC} }I_{DC}+P(I_{DC})}
\end{align}

\begin{figure}[hbt]
\centerline{\includegraphics[width=\linewidth]{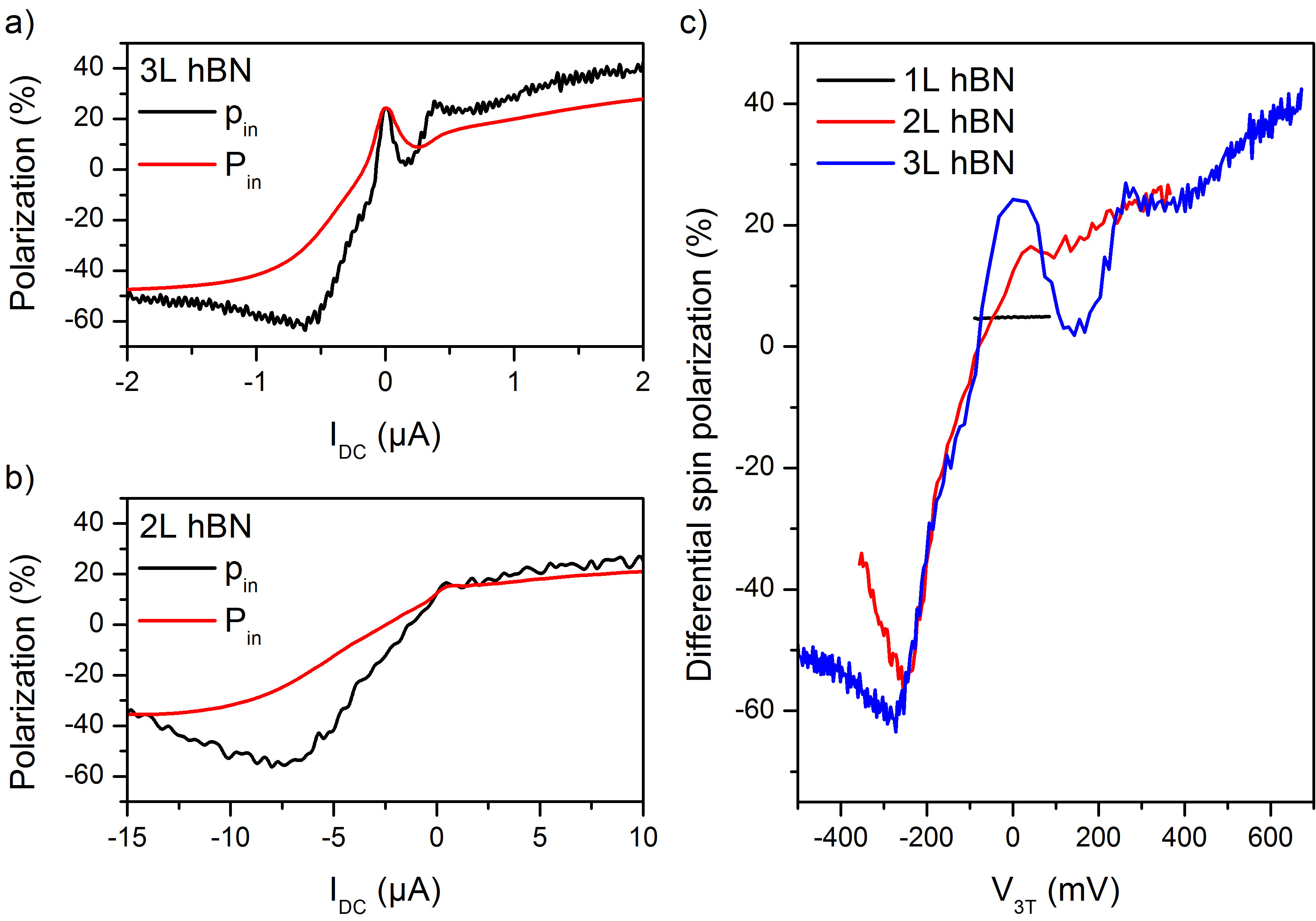}}
\caption{Differential (p$_\mathrm{in}$) and DC (P$_\mathrm{in}$) injector spin polarization of the a) 3L-hBN device using Contact~1 and Contact~5 and b) a 2L-hBN device from Ref.~\cite{Leutenantsmeyer2018YIG}. Note that the numerical integration of p$_\mathrm{in}$ averages the noise out of P$_\mathrm{in}$. c) Comparison of the differential spin polarizations of 1L-, 2L- and 3L-hBN tunnel barriers. The data of 1L-hBN is taken from Ref.~\cite{Gurram2017}.
\label{Figure3_5TL}}
\end{figure}

The results obtained for 3L- and 2L-hBN barriers using this procedure are shown in Fig.~\ref{Figure3_5TL}a and \ref{Figure3_5TL}b. The DC spin polarization of 3L-hBN rises close to 50\%, whereas 2L-hBN yield only up to 35\%. Measurements on vertical tunnel junctions with 1L- and 2L-hBN tunnel barriers reported a spin polarization of $\sim$ 1\% (1L) and 12\% (2L) \cite{Dankert2015,Asshoff2017,Piquemal-Banci2018}. This underlines the potential of cobalt/3L-hBN contacts for highly efficient spin injection into graphene.

The comparison of the differential spin polarization of 1L-, 2L- and 3L-hBN/Co contacts is shown in Fig.~\ref{Figure3_5TL}c. In the case of 1L-hBN, the polarization remains constant ($\sim$ 5\%), mostly independent of the applied V$_\mathrm{3T}$, and clearly below the values of 2L- and 3L-hBN barriers. However, the comparison of 2L- and 3L-hBN yields comparable differential spin polarizations, whereas the electric fields underneath the contacts, which arise from V$_\mathrm{3T}$, change from 1L- to 3L-hBN by a factor of 3. Therefore, local gating underneath the contacts can also be excluded as origin of the bias dependence. The effect of quantum capacitance is discussed in the supplementary information. %The largest electric field resulting from the voltage applied to the 3L-hBN barrier is 0.4~V/nm whereas, for 2L-hBN it is of 0.5~V/nm.

Zollner et al. \cite{Zollner2016} calculated the exchange coupling between cobalt and graphene separated by 1L- to 3L-hBN. Interestingly, they reported a spin splitting of up to 10~meV in when cobalt and graphene are separated by 2L-hBN. For 3L-hBN, this splitting decreases to 18~$\mu$eV. Since we observe very comparable results between 3L-hBN and 2L-hBN, we conclude that proximity induced exchange splitting is most likely not the origin for the DC bias dependent spin injection efficiency in Co/hBN/graphene.

\section{Isotropy of the spin injection efficiency}
By applying a large $\bper \sim$ 1.2~T, we can rotate the cobalt magnetization close to out-of-plane and characterize the spin injection efficiency of 3L-hBN tunnel barrier for out-of-plane spins. This measurement technique was used to determine the spin lifetime anisotropy of graphene \cite{Tombros2008}, which can be also measured using oblique spin precession with lower applied magnetic fields \cite{Raes2016,Raes2017,Leutenantsmeyer2018}. By comparing both results, we can separate the anisotropy of the BLG channel from the anisotropy of the spin injection and detection polarization. 

%To obtain deeper insights on the physics dominating spin injection and the $\vhbn$ dependence of $\mathrm{p_in}$ we measure $\rnl$ up to $\bper =$ 1.2~T, where the cobalt magnetization is rotated out of the plane. 

\begin{figure}[bh]
\centerline{\includegraphics[width=\linewidth]{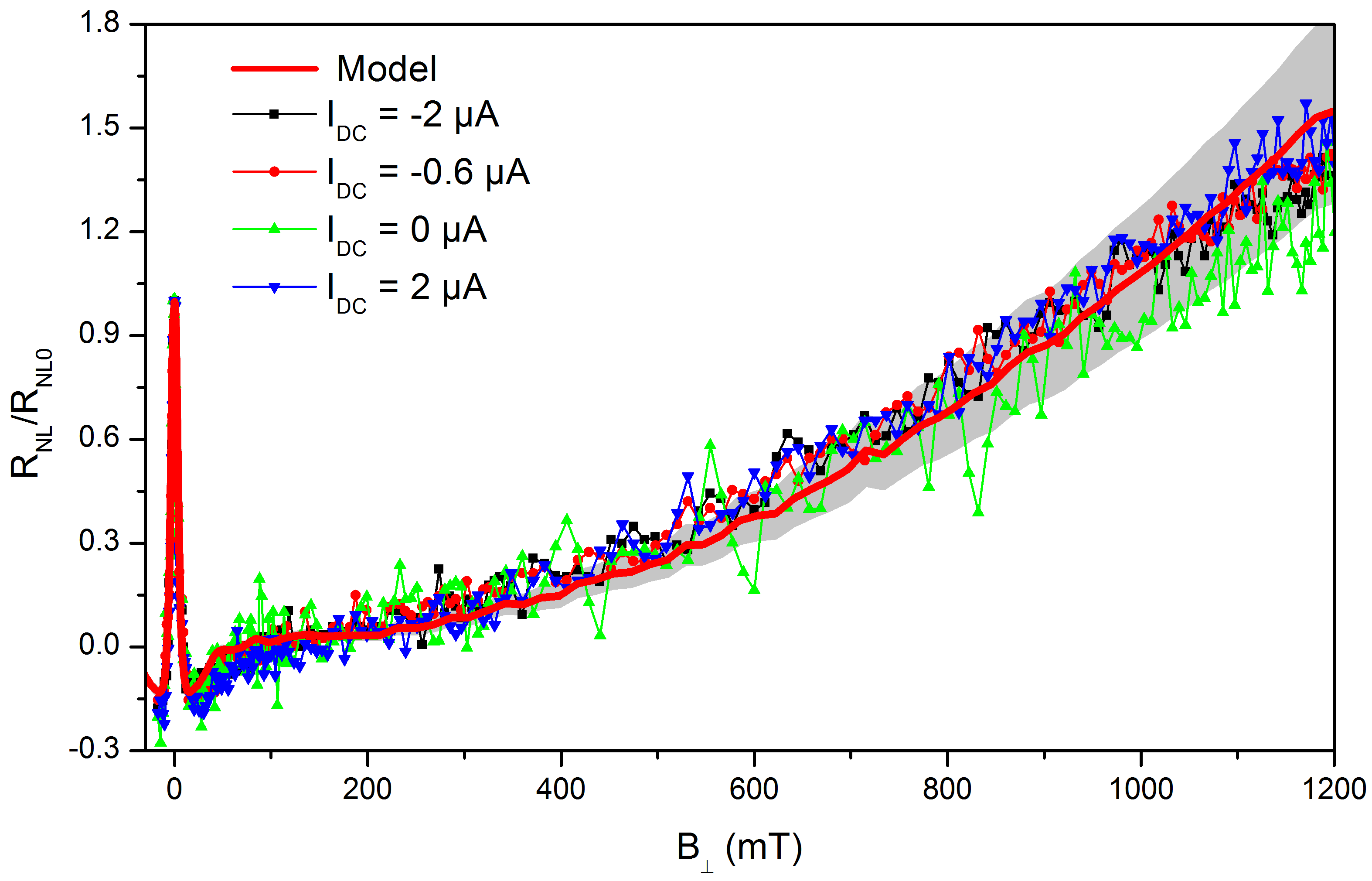}}
\caption{Hanle spin precession curves measured up to $\bper$ = 1.2~T. For comparison, $\rnl$ is normalized to $\rnl$ at $\bper$ = 0 ($\rnlz$). The measurements at different I$_\mathrm{DC}$ are shown as scattered lines, the red solid line is simulated with isotropic spin injection ($\mathrm{\mathrm{p_z}/\mathrm{p_y}} = 1$). %The magnetization rotation from in-plane to out-of-plane is extracted from AMR measurements of cobalt electrodes (see supplementary information). 
\label{Figure4TL}
}
\end{figure}

Fig.~\ref{Figure4TL} shows the Hanle curves measured at a carrier concentration of n = 6 $\times$ 10$^{11}$~cm$^{-2}$, which is the highest density accessible in our device and has been chosen to minimize the effect of magnetoresistance and the spin lifetime anisotropy of the BLG channel. The data is normalized to $\rnlz$ = $\rnl$($\bper$ = 0~T), the gray shaded area is determined by the uncertainty of the extracted spin lifetime anisotropy. The normalized measurements at different I$_\mathrm{DC}$ overlap each other, which indicates that $\mathrm{\mathrm{p_z}/\mathrm{p_y}}$ is independent of $\mathrm{I_{DC}}$. 

We model the spin transport using the Bloch equations for anisotropic spin transport as discussed in Ref.~\cite{Leutenantsmeyer2018}. Additionally, we include the rotation of the contact magnetization, which we extract from anisotropic magnetoresistance measurements, shown in the supplementary information. The good agreement between the experimental data and our model suggests that the spin injection polarization is isotropic, and, hence, $\mathrm{\mathrm{p_z}/\mathrm{p_y}}\approx 1$.

\section{Two terminal DC spin transport measurements up to room temperature}
Lastly, we use the large DC spin polarization of our device to measure spin transport in a local two terminal geometry, which is especially interesting for applications. 
For this experiment we source a DC current ($\idc$) and measure simultaneously the DC voltage V$_\mathrm{DC}$  between Contact~2 and Contact~1. The local, two terminal signal is R$_\mathrm{2T}$ = V$_\mathrm{DC}$/$\idc$, with the spin signal $\Delta\mathrm{R}_\mathrm{2T}$ = $\Delta\mathrm{R}_\mathrm{2T}(\uparrow \uparrow)$ $-$ $\Delta\mathrm{R}_\mathrm{2T}(\uparrow \downarrow)$ is 162~$\Omega$ at $\idc$ = -2~$\mu$A 75~$\Omega$ at $\idc$ = +1~$\mu$A. 

\begin{figure}[tbh]
\centerline{\includegraphics[width=\linewidth]{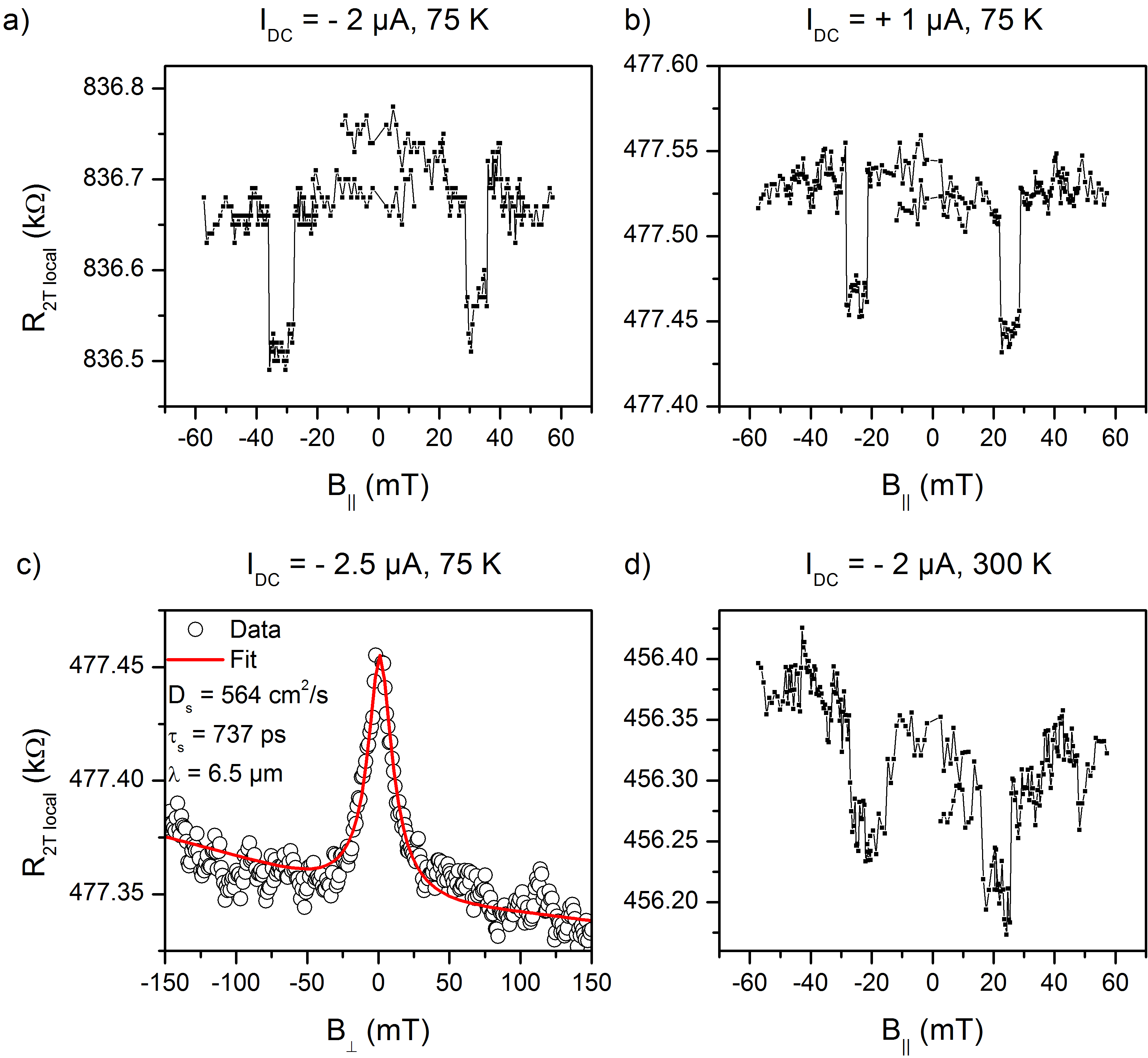}}
\caption{a) Two terminal spin signal measured with $\idc$ = -2~$\mu$A, and b) $\idc$ = +1~$\mu$A. c) Hanle precession data measured at 75~K between Contact~3 and Contact~2 with $\idc$ = -2.5~$\mu$A. d) Room temperature  spin valve measurement between Contact~2 and Contact~1 with I$_\mathrm{DC}$ = -2~$\mu$A.
\label{Figure6}
}
\end{figure}

A measurement of spin precession between Contact~3 and Contact~2 is shown in Fig.~\ref{Figure6}c. We observe a clear Hanle curve and fit the data with $\ts$ = (740 $\pm$ 60)~ps, $\ds$ = (560$ \pm$ 70)~cm$^2$/s and calculate $\lambda$ = 6.5~$\mu$m. Note that the change of these values compared to Table~\ref{Table1_TL} was caused by an exposure of the sample to air. 
Using the spin polarization of the biased contacts and the extracted spin relaxation length, we can calculate the expected local 2T spin valve signal \cite{Gurram2017}:
\begin{equation}
\begin{split}
\Delta\mathrm{R}_\mathrm{2T} &= [ \mathrm{P}_\mathrm{A}(+\mathrm{I}_\mathrm{DC})  \mathrm{p}_\mathrm{B}(-\mathrm{I}_\mathrm{DC}) \\&+ \mathrm{p}_\mathrm{A}(-\mathrm{I}_\mathrm{DC}) \mathrm{P}_\mathrm{B}(+\mathrm{I}_\mathrm{DC})
] \frac{\rsq \lambda}{\mathrm{W}} e^{-\mathrm{L}/\lambda} 
\end{split}
\end{equation}
where the indexes A and B denote both contacts at the bias $\mathrm{I}_\mathrm{DC}$. We calculate using the spin polarization values $\Delta\mathrm{R}_\mathrm{2T}$ = -177~$\Omega$ at $\idc$ = -2~$\mu$A and $\mathrm{R}_\mathrm{2T}$ = -108~$\Omega$ at $\idc$ = +1~$\mu$A, which is in agreement with the measured data in Fig.~\ref{Figure6}a and \ref{Figure6}b of 162~$\Omega$ and 80~$\Omega$.

%(-0,474*0,313-0,424*0,356)*300*6,5/3*exp(-0,6/6,5)
%(-0,21*0,438-0,5934*0,151)*300*6,5/3*exp(-0,6/6,5)

%162~$\Omega$ at $\idc$ = -2~$\mu$A 75~$\Omega$

The measurement of $\Delta\mathrm{R}_\mathrm{2T}$ at room temperature is shown in Fig.~\ref{Figure6}c. $\Delta\mathrm{R}_\mathrm{2T}$ is at room temperature $\sim$ 100~$\Omega$ and clearly present, which indicates no dramatic change of the DC spin polarization with increasing temperature. These results underline the relevance of 3L-hBN barriers for graphene spintronics.

\section{Summary}
In conclusion, we have shown that 3L-hBN tunnel barriers provide a large, tunable spin injection efficiency from cobalt into graphene. The zero bias spin injection polarization is between 20\% and 30\%, and the differential spin injection polarization can increase to -60\% by applying a negative DC bias. The resulting DC spin polarization of up to 50\% allows spin transport measurements in a DC two terminal configuration up to room temperature. 
We study the n dependence of the spin injection polarization and find that it does not depend on n. From a comparison between 3L- and 2L-hBN, we observe that the DC bias dependence scales with the voltage and not the electric field, indicating that local gating is not the dominant mechanism. We also compare the spin injection polarization for in-plane and out-of-plane spins and find that it is isotropic and that $\mathrm{\mathrm{p_z}/\mathrm{p_y}}$ is independent of the applied DC bias. 

During the preparation of this manuscript we became aware of a related work \cite{Zhu2018b}, where also a DC bias dependent spin signal is reported in Co/SrO/graphene heterostructures. Furthermore, the authors also exclude carrier drift as origin. 

%\section{Summary}

\paragraph*{Acknowledgements}
We acknowledge the fruitful discussions with A.A.~Kaverzin and technical support from H.~Adema, J.G.~Holstein, H.M.~de Roosz, T.J.~Schouten, and H.~de Vries. 
This project has received funding from the European Union's Horizon 2020 research and innovation program under the grant agreements 696656 and 785219 (‘Graphene Flagship’ Core~1 and 2), the Marie Curie initial training network ‘Spinograph’ (grant agreement 607904) and the Spinoza Prize awarded to B.J. van Wees by the ‘Netherlands Organization for Scientific Research’ (NWO).
\clearpage
\begin{center}
\textbf{\large Supplementary Information}
\end{center}
\setcounter{equation}{0}
\setcounter{section}{0}
\setcounter{figure}{0}
\setcounter{table}{0}

\renewcommand{\theequation}{S\arabic{equation}}
\renewcommand{\thefigure}{S\arabic{figure}}
\renewcommand{\thesection}{S\arabic{section}}

\section{Fabrication details}
The 3L-hBN/bilayer graphene (BLG)/bottom-hBN stack is fabricated using the scotch tape technique to exfoliate hBN from hBN powder (HQ Graphene) and graphene from HOPG (ZYA grade, HQ Graphene). The materials are stacked using a polycarbonate based dry transfer technique \cite{Zomer2014}. The transfer polymer is removed in chloroform and the sample is annealed for one hour in Ar/H$_2$. PMMA is spun on the sample and contacts are exposed using e-beam lithography. The sample is developed in MIBK:IPA and 65~nm Co and a 5~nm Al capping layer are deposited. The PMMA mask is removed in warm acetone. The sample is bonded on a chip carrier and loaded into a cryostat where the sample space is evacuated below 10$^{-6}$~mbar.

\section{Determination of the unbiased contact spin polarization}
\begin{figure}[htb]
\centerline{\includegraphics[width=\linewidth]{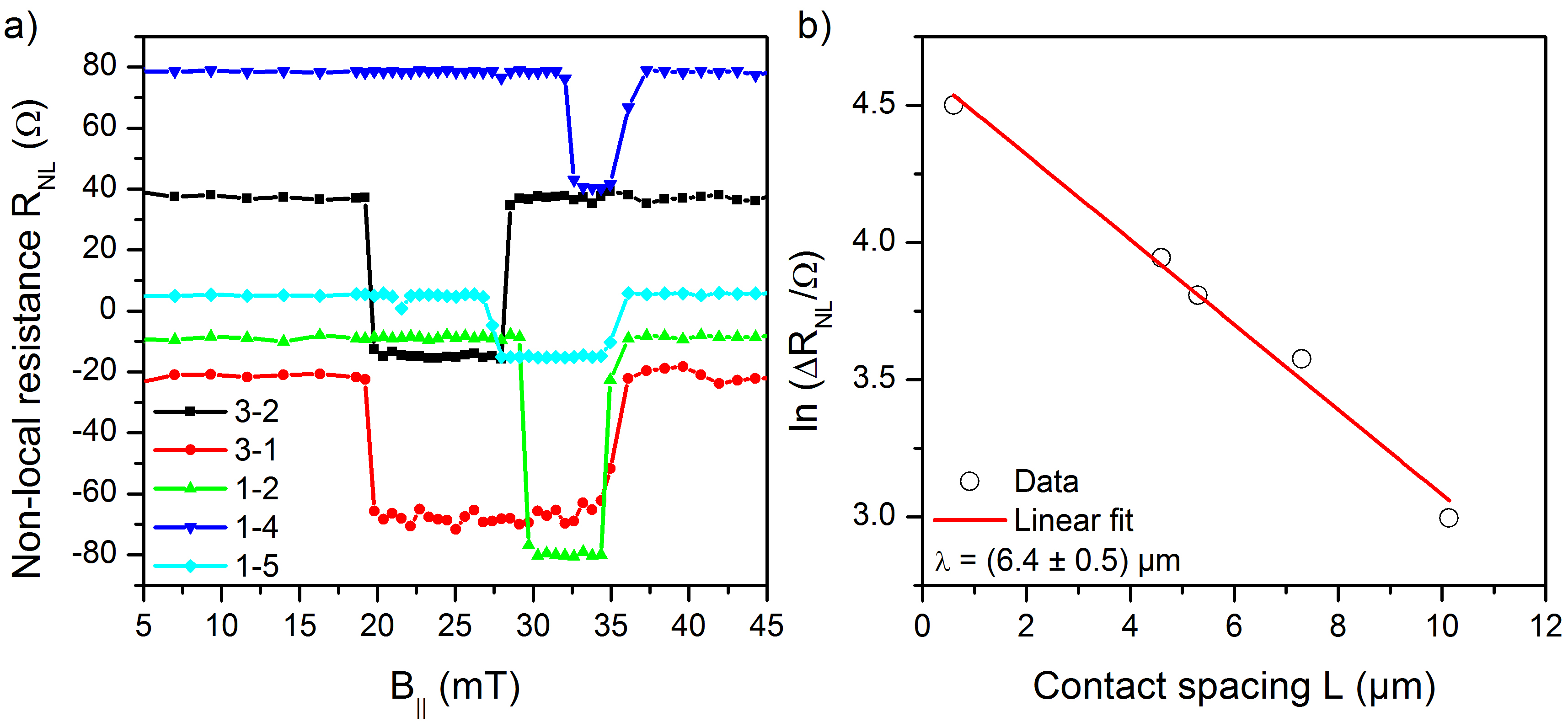}}
\caption{Non-local spin valves for all used injector/detector pairs to determine the spin polarization of each contact.
\label{SI_Spinpol}
}
\end{figure}

Fig.~\ref{SI_Spinpol} shows the non-local spin valve measurement obtained from all different contact combinations. To calculate the unbiased spin polarization of each contact, we apply i$_\mathrm{AC}$ = 50~nA to the injector and obtain the values in Table~\ref{TableSI1}. The measurement is done without any DC bias current and back gate voltage applied, $\mathrm{V}_\mathrm{BG}$ = 0, the corresponding carrier concentration is 4 $\times$ $10^{11}$ cm$^{-2}$.

To calculate the spin polarization of each contact, we use equation~\ref{TLeqnspinpol}:
\begin{align}
\Pa = \frac{\Delta \rnl \mathrm{w}}{\Pb \rsq \lambda} \exp(\mathrm{d}/\lambda)
\label{TLeqnspinpol} % wolfram alpha input: solve {{C=52*3/(B*1000*7.4)*exp(4.6/7.4)},{C=48*3/(A*1000*7.4)*exp(5.3/7.4)},{A=89*3/(B*1000*7.4)*exp(0.6/7.4)},{A=21*3/(D*1000*7.4)*exp(11.1/7.4)},{A=40*3/(E*1000*7.4)*exp(7.3/7.4)}}
%fixed: http://www.wolframalpha.com/input/?i=solve+%7B%7BC%3D52*3%2F(B*1000*5.4)*exp(4.6%2F5.4)%7D,%7BC%3D48*3%2F(A*1000*5.4)*exp(5.3%2F5.4)%7D,%7BA%3D89*3%2F(B*1000*5.4)*exp(0.6%2F5.4)%7D,%7BA%3D21*3%2F(E*1000*5.4)*exp(11.1%2F5.4)%7D,%7BA%3D40*3%2F(D*1000*5.4)*exp(7.3%2F5.4)%7D%7D
\end{align}
where $\Delta \rnl$ is the spin signal extracted from Fig.~\ref{SI_Spinpol}a, w = 3~$\mu$m the width of the BLG and $\rsq$ the square resistance of the BLG. The results are shown in Table~\ref{TableSI1}. %, where the homogeneity of the spin polarization values confirms the outstanding properties of hBN as tunnel barrier for spin injection. 

\begin{table}[htb]
\caption{Measured spin valve signals extracted from the data in Fig.~\ref{SI_Spinpol}. We calculate the unbiased differential spin polarization p of each contact and obtain p$_1$ = 24\%, p$_2$ = 23\%, p$_3$ = 30\%, p$_4$ = 36\%, and p$_5$= 38\% using equation~\ref{TLeqnspinpol}. Note that the larger differential spin polarization values for the larger spacings (Contact~4 and Contact~5) can be explained with the uncertainty in determining the spin relaxation length.}
\begin{ruledtabular}
\begin{tabular}{c c c c }
Injector & Detector & $\Delta\rnl$ ($\ohm$) & d ($\mu$m) \\ \hline
3 & 2 & 52 & 4.6  \\
3 & 1 & 48 & 5.3  \\
1 & 2 & 89 & 0.6  \\
1 & 4 & 40 & 7.3  \\
1 & 5 & 21 & 11.1  \\
\end{tabular}
\end{ruledtabular}
\label{TableSI1}
\end{table}

\section{Extraction of the magnetization rotation through AMR measurements}

\begin{figure}[h]
\centerline{\includegraphics[width=\linewidth]{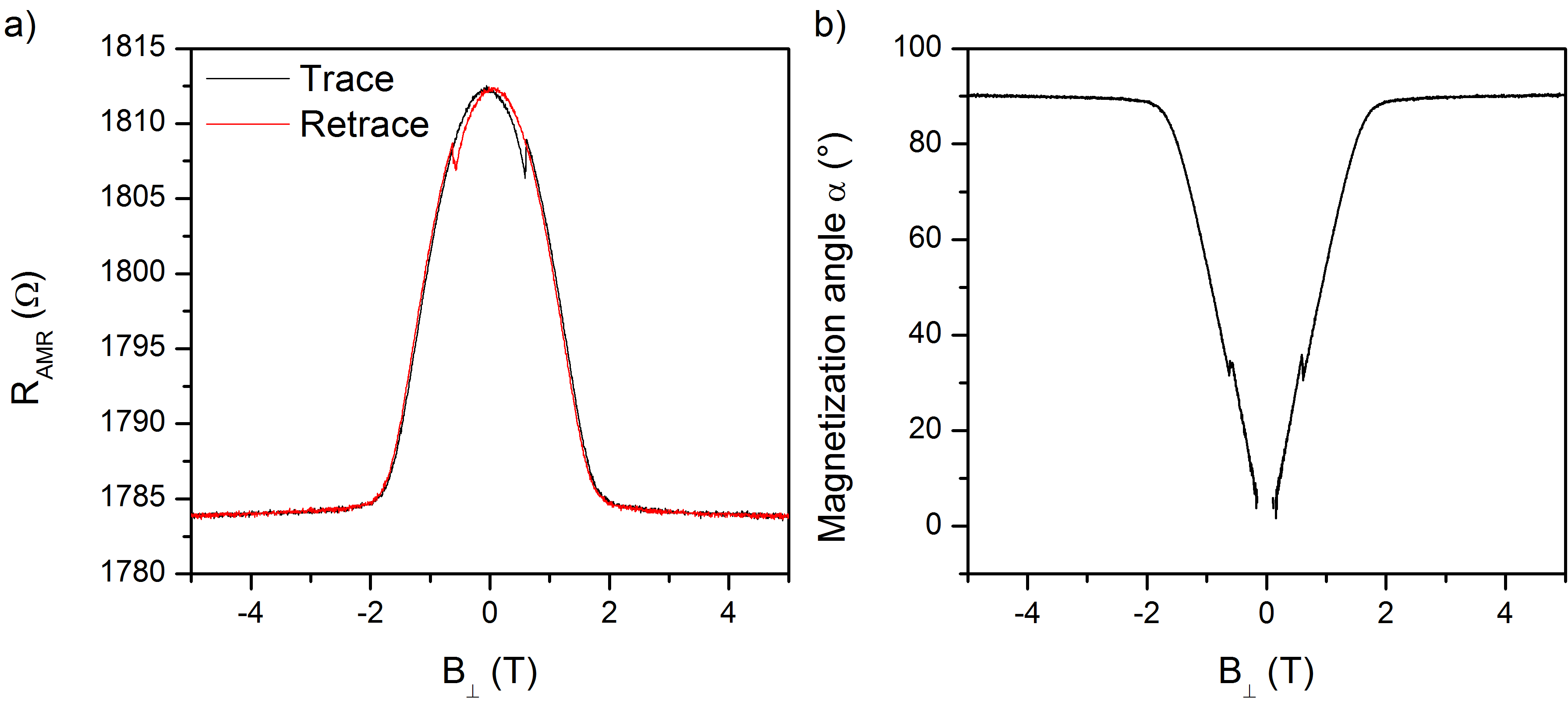}}
\caption{a) AMR measurement of a 65~nm cobalt electrode. b) Calculated magnetization angle under $\bper$.
\label{SI_CobaltAMR}
}
\end{figure}

To accurately model the dependence of $\rnl$ on $\bper$, we measure the anisotropic magnetoresistance (AMR) effect, shown in Fig.~\ref{SI_CobaltAMR}a. The angle of the cobalt magnetization $\alpha$ can be calculated at any given $\bper$ via \cite{Benitez2017}:

\begin{align}
\cos(\alpha(\bper)) = \sqrt{\frac{\mathrm{R}_\mathrm{AMR}(\bper) - \mathrm{R}_\mathrm{AMR}(\bper\,=\,0)}{\mathrm{R}_\mathrm{AMR}(\bper\,=\,2\mathrm{T})-\mathrm{R}_\mathrm{AMR}(\bper\,=\,0)}}
\end{align}
The calculated magnetization angle $\alpha(\bper)$ is shown in Fig.~\ref{SI_CobaltAMR}b and used to model the spin precession curves in the main text. 

\section{Full set of spin transport parameters}
Table~\ref{TableSI2} contains an overview of the full set of spin transport measurements using Contact~1 as injector and Contact~5 as detector. For each applied gate voltage, all spin transport parameters are within the experimental uncertainty, implying the independence of $\ds$ and $\ts$ on $\mathrm{I}_\mathrm{DC}$.

\begin{table}[htb]
\caption{Basic spin and charge transport parameters measured by Hanle spin precession using Contact~1 and Contact~5.}
\begin{ruledtabular}
\begin{tabular}{c c c c c c}
 $\mathrm{I}_\mathrm{DC}$& $\mathrm{V}_\mathrm{BG}$&$\rsq$& $\ds$  & $\ts$ & $\lambda$ \\
($\mu$A) & (V) & ($\ohm$) &  (cm$^2$/s) & (ns) &($\mu$m)\\ \hline
-2	&-2	& 1600 &112	 $\pm$ 27	&		1.5 $\pm$ 0.3&	4.1	$\pm$ 1.8 \\
-0.6	&-2	& 1600 &114	 $\pm$ 24	& 1.4 $\pm$	0.3&	4.0	$\pm$ 1.6 \\
0	&-2	&1600 &138 $\pm$ 	43	& 		1.6 $\pm$	0.5&	4.8	$\pm$	2.8 \\
2&	-2&1600 &	86	 $\pm$ 21	& 		1.0 $\pm$	0.3&	2.9 $\pm$	1.3 \\ \hline
-2&	-1	&1400 &133	 $\pm$ 22	& 	1.7 $\pm$	0.3&	4.8	$\pm$	1.5 \\
-0.6&	-1&1400 &	140	 $\pm$ 20	& 	1.7 $\pm$	0.3&	4.9	$\pm$	1.3 \\
0&	-1&1400 &	160	 $\pm$ 40	& 		1.8 $\pm$	0.4&	5.4	$\pm$	2.5 \\
2&	-1	&1400 &137	 $\pm$ 22	 &	1.8 $\pm$	0.3&	5.0	$\pm$	1.5 \\ \hline
-2&	0	&900 &202	 $\pm$ 26&				2.0 $\pm$	0.3&	6.4	$\pm$	1.6 \\
-0.6&	0	&900 &176 $\pm$ 	21	&		1.7 $\pm$	0.2&	5.5	$\pm$	1.2 \\
0&	0&900 &170 $\pm$ 	24	&				1.7 $\pm$	0.3&	5.4	$\pm$	1.5 \\
2&	0	&900 &174	 $\pm$ 24	&				1.9 $\pm$	0.3&	5.8	$\pm$	1.5 \\ \hline
-2&	1&	750 &226 $\pm$ 	24	&			2.2 $\pm$	0.2&	7.1	$\pm$	1.4 \\
-0.6&	1	&750 &230	 $\pm$ 25	&			1.8 $\pm$	0.2&	6.5	$\pm$	1.3 \\
0&	1	&750 &214 $\pm$ 	27&				1.8 $\pm$	0.2&	6.1	$\pm$	1.5 \\
2&1&	750 &222 $\pm$ 25	&					2.1 $\pm$ 0.2&	6.8	$\pm$	1.4 \\
\end{tabular}
\end{ruledtabular}
\label{TableSI2}
\end{table}

\section{Quantum capacitance correction to bias-induced gating}
A gate voltage does not only apply an electric field to the graphene channel but also tunes the Fermi energy (E$_\mathrm{F}$). This effect is called quantum capacitance correction and becomes relevant when the geometrical capacitance of the gate is very high, or the density of states of the channel is small. %In our case, when determining the carrier density induced by the biased 3L-hBN contacts on the graphene, we need to take this correction into account. 
The quantum capacitance correction is calculated via \cite{Braga2012}:

\begin{equation} \label{TLEqCap}
\mathrm{e} \Delta \mathrm{V}_\mathrm{c} = \Delta \mathrm{E}_\mathrm{F} + \frac{\mathrm{e}^2 \Delta \mathrm{n} \mathrm{t}_\mathrm{hBN}}{\epsilon_0 \epsilon_r}
\end{equation}
where V$_\mathrm{c}$ denotes the voltage applied to the contact, $\mathrm{t_{hBN}}$ the hBN tunnel barrier thickness, $\mathrm{e}$ the electron charge, $\epsilon_0$ the vacuum permittivity, and $\epsilon_r$ the relative permittivity of hBN. %The right term describes the quantum capacitance correction

\begin{table}[b]
\caption{Calculation of the quantum capacitance corrections. The change in n induced by the bias applied to the contacts $\mathrm{n_{corr}}$ is determined using Equation~\ref{TLEqCap}. The classical gating $\mathrm{n_{geo}}$ is shown for comparison. The relative permittivities are taken from Ref.~\cite{Laturia2018}.}
\begin{ruledtabular}
\begin{tabular}{c c c c c}
 $\mathrm{V}_\mathrm{C}$& $\mathrm{\epsilon}_\mathrm{r}$&$\mathrm{t_{hBN}}$& $\mathrm{n_{corr}}$&$\mathrm{n_{geo}}$   \\
(mV) &  & (nm) &  (cm$^{-2}$) & (cm$^{-2}$) \\ \hline
300 & 3.52 & 1.2 (3L) & 3.35$\times10^{12}$&4.86$\times10^{12}$ \\
300 & 3.44 & 0.7 (2L) & 4.78$\times10^{12}$&8.15$\times10^{12}$ \\
\end{tabular}
\end{ruledtabular}
\label{TableSI3TL}
\end{table} 

The Fermi energy in the conduction band of BLG is determined by \cite{Mccann2013}:
\begin{equation}
\mathrm{E_F=-\frac{\gamma_1}{2}+\frac{\sqrt{\gamma_1^2+4n\pi \hbar^2v_F^2}}{2}}\label{TLEqEf}
\end{equation} 
where $\gamma_1$ is the interlayer hopping constant, %$\mathrm{\alpha=(\pi \hbar^2v_F^2)^{-1}}$, 
$\hbar$ the reduced Plank constant, and $\mathrm{v_F = 10^6}$~m/s the Fermi velocity in graphene. 

Using Equation~\ref{TLEqCap} and Equation~\ref{TLEqEf}, we calculate the carrier density for a DC bias of 300~mV in Table~\ref{TableSI3TL}, assuming that the charge neutrality point lies at zero DC bias. We find that the quantum capacitance can have a significant effect on the carrier density $\mathrm{n_{corr}}$ compared to  classical gating $\mathrm{n_{geo}=\epsilon_0\epsilon_rV_c/(et_{hBN})}$.

In conclusion, we find a substantial quantum capacitance correction. However, even with the quantum correction applied, the difference in the carrier concentration of 2L- and 3L-hBN is $\sim$ 30\%. Consequently, we can still exclude local gating as origin of the DC bias dependence. %, since  would be the reason for the bias dependence, we would expect a significant difference in the $\mathrm{V_{C}}$ required to achieve the maximal differential spin polarizations. 

\bibliography{AllReferences}% Produces the bibliography via BibTeX.

\end{document}